\documentclass{appolb}
\usepackage{graphicx}
\usepackage{amsmath,amssymb}
\newcommand{\calA}{\mathcal{A}}
\newcommand{\calB}{\mathcal{B}}
\newcommand{\calD}{\mathcal{D}}
\newcommand{\calE}{\mathcal{E}}
\newcommand{\calF}{\mathcal{F}}
\newcommand{\calO}{\mathcal{O}}
\newcommand{\calP}{\mathcal{P}}
\newcommand{\calW}{\mathcal{W}}
\newcommand{\xt}{\boldsymbol{x}_\perp}
\newcommand{\yt}{\boldsymbol{y}_\perp}
\newcommand{\br}{\boldsymbol{r}}
\newcommand{\bdel}{\boldsymbol{\partial}_\perp}
\newcommand{\rhot}{\rho_{\mathrm{t}}}
\newcommand{\rhop}{\rho_{\mathrm{p}}}
\newcommand{\rmd}{\mathrm{d}}
\newcommand{\rme}{\mathrm{e}}
\newcommand{\rmi}{\mathrm{i}}
\newcommand{\Nc}{N_{\mathrm{c}}}
\newcommand{\tr}{\mathrm{tr}}

\begin{document}
\title{Evolving Glasma and Kolmogorov Spectrum%
\thanks{Presented at the 51st Cracow School of Theoretical Physics}%
}
\author{Kenji Fukushima
\address{Department of Physics, Keio University,
         Kanagawa 223-8522, Japan}
}
\maketitle
\begin{abstract}
 We present a pedagogical introduction to the theoretical framework of
 the Color Glass Condensate (CGC) and the McLerran-Venugopalan (MV)
 model.  We discuss the application of the MV model to describe the
 early-time dynamics of the relativistic heavy-ion collision.  Without
 longitudinal fluctuations the classical time evolution maintains
 boost invariance, while an instability develops once fluctuations
 that break boost invariance are included.  We show that this
 ``Glasma instability'' enhances rapidity-dependent variations as long
 as self-interactions among unstable modes stay weak and the system
 resides in the linear regime.  Eventually the amplitude of unstable
 modes becomes so large that the growth of instability gets saturated.
 In this non-linear regime the numerical simulations of the Glasma
 lead to turbulent energy flow from low-frequency modes to
 higher-frequency modes, which results in a characteristic power-law
 spectrum.  The power found in numerical simulation of the expanding
 Glasma system turns out to be consistent with Kolmogorov's $-5/3$
 scaling.
\end{abstract}

\section{Introduction}

  Relativistic heavy-ion collision experiments have aimed to create a
new state of matter out of color-deconfined particles, i.e.\ a
quark-gluon plasma (QGP) in extreme environments in the laboratory.
Presumably it was only up to $\sim10^{-6}{\;\rm s}$ after the Big Bang
when the Early Universe was still hot enough to realize the QGP in
nature.  Experimental data on intrinsic properties of the QGP
suggest that this new state of QCD matter found in the heavy-ion
collision is not a weakly-coupled plasma but rather a strongly-coupled
fluid.  The hydrodynamic description of the time evolution has
successfully reproduced the measured particle distributions, in
particular, the azimuthal distribution of emitted particles in non-central
collisions.  The great success of the hydrodynamic model
is a strong evidence for thermalization.  After thermal equilibrium is
achieved, the time evolution of the QGP is somehow under theoretical
control, though some uncertainties remain in the determination of the
equation of state and the implementation of dissipative effects.

  In contrast to the hydrodynamic regime after thermalization, our
understanding is still quite limited about the early-time dynamics
toward thermalization.  This kind of problem is in general one of the
most difficult physics challenges.  One would naturally anticipate
that the system may form a turbulent fluid during transient stages
right after the collision.  In fact, turbulence is a common phenomenon
in our daily life whenever the Reynolds number exceeds a certain
threshold.  Nevertheless, it poses a very difficult theory problem even
today.  It is widely known that a renowned physicist, Richard Feynman
described turbulence as ``the last great unsolved problem of classical
physics'' and it is indeed so also in the context of the QGP study.

  A good news is that we already have a powerful theoretical tool to
investigate the very early-time dynamics of the high-energy heavy-ion
collision.  One may well think at a first glance that any microscopic
description of nucleus-nucleus collisions is too complicated to handle
in terms of the QCD first principle.  This is truly so unless the
collisional energy is sufficiently high.  Actually microscopic
information of nucleus before collision is far from simple on its own.
The high-energy limit, however, allows for drastic simplification that
makes the calculation feasible.  In the Regge limit, precisely
speaking, the strong interactions exhibit totally different
characteristics from low-energy hadron physics.  In this particular
case the c.m.\ energy scale $s$ is infinitely larger than other energy
scales such as the transferred momentum squared $t$.  Then, although
the strong coupling constant $\alpha_s$ is small due to asymptotic
freedom with large $s$, a resummation is required for a series of
non-small terms $(\alpha_s \ln x^{-1})^n$ where Bjorken's
$x\sim t/s$.  In terms of the Feynman diagram this resummation
represents quantum processes to emit softer gluons successively.
Intuitively, as one goes to larger $s$ and thus goes to
$x-\delta x$ down from $x$, one should consider more radiated gluons
within a bin of $x$ to $x-\delta x$.  Naturally the wave function of
nucleus is $x$-dependent and we should expect more and more gluons
inside at smaller $x$.

  Eventually it should be the most suitable to treat gluons as
coherent fields rather than particles once the gluon density is high
enough.  This is reminiscent of photons in the Weizs\"{a}cker-Williams
approximation.  In the first approximation in the high-energy limit of
QCD, therefore, the \textit{classical} fields are appropriate
ingredients in theoretical computations, which is as a consequence of
\textit{quantum} radiations.  Such a classical description of
high-energy QCD is called the Color Glass Condensate
(CGC)~\cite{review}.  In this way, the initial condition for the
relativistic heavy-ion collision should be formulated by means of the
CGC theory.  As long as the gluon distribution function stays large,
the CGC picture holds, and it is finally superseded by a particle
picture of plasma.  Some unnamed physical state between the CGC and
the QGP was given a name of Glasma as a mixture of ``glass'' and
``plasma''~\cite{Lappi:2006fp}.  The Glasma time evolution thus
follows the classical equations of motion and hopefully it should be
transformed smoothly into a hydrodynamic regime.  In this sense the
Glasma should play a central role to figure out what the initial
condition for the hydrodynamic model should look like.

  Because QCD or the Yang-Mills theory involves gluonic
self-interactions, it is generally hard to find an analytical solution
of the classical equations of motion except for some simple
situations.  Hence, one has to resort to numerical methods to go into
quantitative estimates, and the numerical implementation has been
established.  In view of numerical outputs, however, there is no
indication seen in the Glasma evolution toward thermalization.  If all
quantum fluctuations are completely frozen in particular, the
classical dynamics respects boost invariance.  The coordinate rapidity
$\eta$ is simply shifted under a boost on the system, that means that
the boost-invariant system is insensitive to $\eta$.  The pressure is
then highly anisotropic depending on the beam-axis direction or on the
transverse plane.  This makes a sharp contrast to thermalized matter
in which the pressure is isotropic.

  Quantum (or structural -- see Sec.~\ref{sec:inst}) fluctuations
including $\eta$-dependence break boost invariance explicitly.
Interestingly enough, it was discovered that a small modulation along
the longitudinal direction grows up exponentially as a function of
time and $\eta$-dependent modes show instability behavior, which is
sometimes referred to as the ``Glasma
instability''~\cite{Romatschke:2005pm}.  It is of paramount importance
to understand the nature of the Glasma instability to fill in a
missing link between the CGC initial state and the initial condition
for hydrodynamics.  There are several theoretical attempts to account
for qualitative features of the Glasma
instability~\cite{Fukushima:2007ja,Fujii:2008dd,Iwazaki:2008xi,%
Dusling:2010rm,Dusling:2011rz,Epelbaum:2011pc}.
Here, instead of doing so, we will think about subsequent phenomena at
later stages;  turbulence may be formed by instability growth, and
then it is sensible to anticipate the decay of turbulence and the
associated scaling law in the energy
spectrum~\cite{Fukushima:2011nq}.

  This article contains lectures on the basic facts of the MV model
for those who are not necessarily familiar with QCD at high energy.
In Sec.~\ref{sec:Eikonal} a stationary-point approximation on the
functional integration is introduced, which leads to a classical
treatment of the high-energy QCD problems.  The classical equations of
motion in the pure Yang-Mills theory are further discussed in
Sec.~\ref{sec:MV}.  Then, some numerical results for the
$\eta$-independent case are presented next in
Sec.~\ref{sec:numerical}, and those for the case with $\eta$-dependent
fluctuations in Sec.~\ref{sec:inst}.  Some evidence is presented for
the realization of the power-law scaling in the energy spectrum in the
non-linear regime where the instability stops.
Section~\ref{sec:outlook} is devoted to outlooks.

\section{Scattering amplitude and the Eikonal approximation}
\label{sec:Eikonal}

  We will see the essence of the Eikonal approximation and the
scattering amplitudes of our interest in QCD physics.  We will first
consider the case of light projectile and dense target, and proceed to
the case of dense projectile and dense target next.

\subsection{Eikonal approximation}

\begin{figure}
\begin{center}
\includegraphics[width=0.5\textwidth]{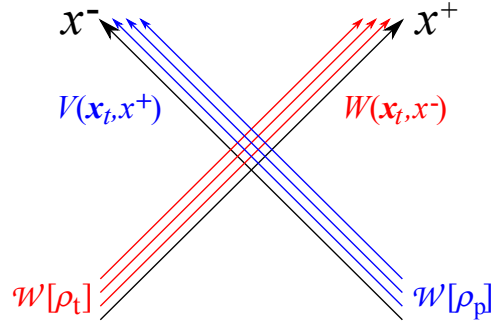}
\end{center}
\caption{Schematic figure to show scattering between the target
  partons along the $x^+$-axis and the projectile partons along the
  $x^-$-axis.}
\label{fig:scattering}
\end{figure}

  Before addressing QCD application, we shall first consider a
scattering problem in non-relativistic Quantum Mechanics.  If we want
to solve a problem of potential scattering, we should treat the
Schr\"{o}dinger equation,
\begin{equation}
 \biggl[-\frac{\boldsymbol{\partial}^2}{2m} + V(r) \biggr]\psi(\br)
  = E\,\psi(\br)
\end{equation}
with the following boundary condition,
\begin{equation}
 \psi(\br) = \rme^{\rmi k z} + f(\Omega)\frac{\rme^{\rmi k r}}{r}
\end{equation}
at large distance ($r\to \infty$).  The term involving $f(\Omega)$
represents the scattered wave and one can obtain the scattering
amplitude from $|f(\Omega|^2$.  The lowest-order estimate immediately
gives an expression in the Born approximation.  In the high-energy
limit, however, the scattering angle is small and the incident and the
scattered waves interfere strongly there.  In this situation the
following Ansatz is more convenient,
\begin{equation}
 \psi(\br) = \rme^{\rmi k z}\,\hat{\psi}(\br) \;,
\end{equation}
which is called the Eikonal approximation in analogy with the
terminology in Optics.  The wave length of incident wave,
$\lambda=2\pi/k$, is shorter than the potential range when $k$ is
large enough, and the differential equation for $\hat{\psi}(\br)$ is
\begin{equation}
 \biggl[ v \hat{p}_z - \frac{\boldsymbol{\partial}^2}{2m}
  + V(r)\biggr] \hat{\psi}(\br) = 0
\end{equation}
with $v=\hbar k/m$.  Then, in the high-energy limit, the second term
is negligible as compared to the first term, which is easily
integrated out to give,
\begin{equation}
 \hat{\psi}(\br) = \exp\biggl[ -\frac{\rmi}{v}\int_{-\infty}^z 
  \rmd z'\;V(x,y,z')\biggr] \;.
\label{eq:eikonal}
\end{equation}
In the gauge theory $\rmd z'V(z')$ is replaced by
$\rmd z'A_{z'}(z')$.  Let us consider the scattering of the target
particles moving at the speed of light in the positive-$z$ direction
and the projectile particles in the negative-$z$ direction.  Then, in
general, in the Eikonal approximation, the scattering matrix is
\begin{equation}
 S \sim \biggl\langle \sum_{\{\rhot\}} \calW_x[\rhot]
  \prod_{\{\rho_t\}} W \; \sum_{\{\rhop\}} \calW_{x'}[\rhop]
  \prod_{\{\rho_p\}} V \biggr\rangle
\label{eq:s}
\end{equation}
with the Wilson lines corresponding to the Eikonal
phase~(\ref{eq:eikonal}),
\begin{equation}
 \begin{split}
 & W(\xt,x^-) = \calP\exp\biggl[ \,\rmi g \int \rmd x^+
  A^-(x^+,x^-,\xt) \biggr] \;, \\
 & V(\xt,x^+) = \calP\exp\biggl[ \,\rmi g \int \rmd x^-
  A^+(x^+,x^-,\xt) \biggr]
 \end{split}
\end{equation}
in the light-cone coordinates; $x^+=(z+t)/\sqrt{2}$ and
$x^-=(z-t)/\sqrt{2}$, as sketched in Fig.~\ref{fig:scattering}.
The momenta conjugate to $x^+$ and $x^-$ are
the energy $p^-=(E-p^z)/\sqrt{2}$ and the longitudinal momentum
$p^+=(E+p^z)/\sqrt{2}$.  In Eq.~(\ref{eq:s}) the weights,
$\calW_x[\rhot]$ and $\calW_{x'}[\rhop]$, represent the wave functions
of the target and the projectile at the Bjorken variables $x$ and
$x'$, respectively.  Since the hard particles ($p^+>x P^+$ with $P^+$
being the total longitudinal momentum of the target) are included in
the wave function, the functional integration in
$\langle\cdots\rangle$ should contain the softer gauge fields with
$p^+<xP^+$.  In the following subsections let us discuss how to make
an approximation on this scattering amplitude.

\subsection{Light projectile and dense target}

  For the simplest example let us take the projectile as a
color-dipole, i.e.\ the scattering amplitude is then,
\begin{align}
 S_{\text{dipole}} &\sim
  \raisebox{-7pt}{\includegraphics[width=2cm]{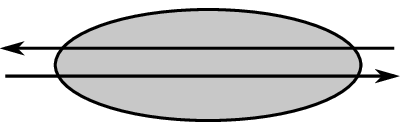}} \notag\\
 &= \langle\!\langle V(\xt) V^\dagger(\yt)
  \rangle\!\rangle_{\rhot} \notag\\
 &= \biggl\langle \sum_{\{\rhot\}} \calW_x[\rhot]\;
  V(\xt) V^\dagger(\yt) \biggr\rangle \notag\\
 &= \sum_{\{\rhot\}} \calW_x[\rhot] \int_{p^+<xP^+} \!\!\!\!\!\!\!\!
  \calD A\; V(\xt)V^\dagger(\yt)\; \rme^{\rmi S_{\rm YM}[A]
  + \rmi S_{\text{source}}[\rhot,W]} \;.
\label{eq:dipole}
\end{align}
The product of $W$'s can be re-expressed on the
exponential~\cite{Jalilian-Marian:2000ad,Fukushima:2005kk} as
\begin{equation}
 S_{\text{source}}[\rhot,W] = \frac{\rmi}{g\Nc}\int\rmd^4 x\,
  \tr\bigl[\rhot \ln W] \sim -\int \rmd^4 x\, \rhot^a A_a^- \,,
\end{equation}
where the last expressions is an approximation valid for sufficiently
large $\rhot$.  Then, with large $\rhot$, the functional integration
in Eq.~(\ref{eq:dipole}) can be estimated by means of the stationary
approximation at the solution of
\begin{equation}
 \frac{\delta S_{\rm YM}}{\delta A_a^\mu}\biggr|_{A=\calA}
  = \delta^{\mu-} \rhot \;.
\label{eq:eom}
\end{equation}
The solution of the above classical equations of motion thus
represents the contribution from soft gluons with $p^+<xP^+$.  Then,
finally, the dipole scattering amplitude is
\begin{equation}
 S_{\text{dipole}} \simeq \sum_{\{\rhot\}}\calW_x[\rhot]\,
  V(\xt) V^\dagger(\yt) \Bigr|_{A=\calA[\rhot]} \;.
\end{equation}
This expression is easily generalized to an arbitrary operator
$\calO[A]$ and
\begin{equation}
 \langle\!\langle \calO[A] \rangle\!\rangle_{\rhot}
  \simeq \sum_{\{\rhot\}}\calW_x[\rhot]\,\calO[\calA[\rhot]] \;.
\label{eq:one-source}
\end{equation}
Once the $x$ dependence in $\calW_x[\rhot]$ is
known~\cite{Iancu:2000hn,Ferreiro:2001qy}, small-$x$ evolution is
deduced for $\langle\!\langle\calO[A]\rangle\!\rangle_{\rhot}$ in
general and in this way the BFKL equation up to the quadratic-order of
$\rhot$, and besides, the BK~\cite{Kovchegov:1999yj} and JIMWLK
equations including full-order of $\rhot$ are derived.

\subsection{Dense projectile and dense target}

  The discussions so far are quite generic but the above
stationary-point approximation needs slight modifications when not
only the target parton density $\rhot$ but also the projectile $\rhop$
is large as in the situation of the relativistic heavy-ion collision.
Then, the scattering amplitude reads
\begin{equation}
 S_{\text{dense-dense}} = \sum_{\{\rhot,\rhop\}} \calW_x[\rhot]\,
  \calW_{x'}[\rhop]\, \int\calD A\, \rme^{\rmi S_{YM}[A]
  +\rmi S_{\text{source}}[\rhot,W;\,\rhop,V]}
\end{equation}
with the source action given approximately as
\begin{equation}
 S_{\text{source}}[\rhot,W;\,\rhop,V] \sim -\int\rmd^4 x\,
  \bigl( \rhot^a A_a^- + \rhop^a A_a^+ \bigr) \;.
\end{equation}
This time, the stationary-point is shifted also by the effect of the
presence of $\rhop$ and it is determined by the classical equations of
motion with two sources,
\begin{equation}
 \frac{\delta S_{\rm YM}}{\delta A_a^\mu}\Bigr|_{A=\calA}
  = \delta^{\mu-}\rhot^a
  +\delta^{\mu+}\rhop^a \;.
\label{eq:eom2}
\end{equation}
Using the solution $\calA$ of the above classical equations of motion,
one can obtain the general formula similar to previous
Eq.~(\ref{eq:one-source}) as
\begin{equation}
 \langle\!\langle \calO[A] \rangle\!\rangle_{\rhot,\rhop}
  \simeq \sum_{\{\rhot,\rhop\}} \calW_x[\rhot]\,\calW_{x'}[\rhop]\,
  \calO[\calA[\rhot,\rhop]] \;.
\label{eq:dense-dense}
\end{equation}
In what follows we discuss how to solve this above
Eq.~(\ref{eq:dense-dense}) to investigate the early-time dynamics of
the heavy-ion collision.

Then, there are two ingredients necessary to estimate physical
observables using Eq.~(\ref{eq:dense-dense}).  One is the solution of
Eq.~(\ref{eq:eom2}), which is impossible to find in an analytical way
unfortunately.  The other is to figure out the wave functions
$\calW_x[\rhot]$ and $\calW_{x'}[\rhop]$, which is again impossible to
do so by solving QCD exactly.  Given an initial condition at a certain
$x$, in principle, the evolution equation such as the JIMWLK equation
leads to the wave function at any $x$.  The theoretical framework of
such a description of scattering processes with non-linearity of
abundant gluons is called the Color Glass Condensate (CGC).  The CGC
theory is not a phenomenological model but an extension of
conventional perturbative QCD with resummation in a form of background
fields.  In any case, however, the initial condition for small-$x$
evolution is necessary for actual computations.  This part needs an
Ansatz as explained in the next section.

\section{Equations of motion and the MV model}
\label{sec:MV}

Here we will see that the analytical solution is written down for
one-source problem.  Although it is impossible to give an analytical
formula for the solution of two-source problem, the initial condition
on the light-cone can be specified.  Also we will introduce a Gaussian
approximation for the wave functions that defines the MV model.

\subsection{One-source problem}

Let us first consider how to solve the equations of
motion~(\ref{eq:eom}) for the light-dense scattering.  It is actually
easy to find a special solution in the same way as the classical
electromagnetism.  The important point is that the source
$\rhot(\xt,x^-)$ is independent of $x^+$ because of time dilation of
particles moving at the speed of light\footnote{Even though the time
  dependence is frozen at the speed of light, non-commutativity of
  color charges needs distinction by the order of interactions along
  the $x^+$-axis, which introduces a label that plays the role of
  time.  Such ``$x^+$-dependence'' is dropped off in the large $\Nc$
  limit only.}.  Therefore, with an assumption of
$\calA^+\neq0$ and $\calA^-=\calA^i=0$, the problem is reduced to an
Abelian one and Eq.~(\ref{eq:eom}) amounts to the standard Poisson
equation.  The solution of the static potential therefore reads,
\begin{equation}
 -\bdel^2 \calA^+ = \rhot(\xt,x^-) \quad\Rightarrow\quad\;
  \calA^+ = -\frac{1}{\bdel^2}\rhot(\xt,x^-) \;.
\end{equation}
In later discussions it is more convenient to adopt the light-cone
gauge, $\calA^+=0$, which is achieved by a gauge rotation by
$V^\dagger$ that solves
$[\calA^+-(\rmi g)^{-1}\partial^+]V^\dagger=0$.  That is,
\begin{align}
 V^\dagger(\xt,x^-) &= \calP\exp\biggl[\,\rmi g \int_{-\infty}^{x^-}
  \rmd z^- \calA^+(\xt,z^-) \biggr] \notag\\
 &= \calP\exp\biggl[\,-\rmi g \int_{-\infty}^{x^-} \rmd z^-
  \frac{1}{\bdel^2}\rhot(\xt,z^-)\biggr] \;.
\label{eq:V}
\end{align}
After the gauge transformation by $V^\dagger$, only the transverse
components are non-vanishing,
\begin{equation}
 \alpha_i^{\rm (t)}=\calA_i=-\frac{1}{\rmi g}V\partial_i V^\dagger \;,
\end{equation}
and $\calA^+=\calA^-=0$.  In the same way, for the projectile moving
in the opposite direction to the target, the equations of motion have
a solution as
\begin{equation}
 \alpha_i^{\rm (p)} = -\frac{1}{\rmi g}W\partial_i W^\dagger
\end{equation}
with
\begin{equation}
 W^\dagger(\xt,x^+) = \calP\exp\biggl[\,-\rmi g
  \int_{-\infty}^{x^+} \rmd z^+ \frac{1}{\bdel^2}
  \rhop(\xt,z^+)\biggr]
\label{eq:W}
\end{equation}
in $\calA^-=0$ gauge.  In this manner the one-source problem is
readily solvable.  It is, however, impossible to solve the two-source
problem in Eq.~(\ref{eq:eom2}) as simply as above.

\subsection{Two-source problem}

  In the presence of two sources it is most suitable to make use of
the Bjorken coordinates that represent an expanding system.  The time
and the longitudinal variables are replaced by the proper time and the
space-time rapidity, respectively, as
\begin{equation}
 \tau = \sqrt{2x^+ x^-} = \sqrt{t^2-z^2}\;,\qquad
 \eta = \frac{1}{2}\ln\frac{x^+}{x^-}
  = \frac{1}{2}\ln\biggl(\frac{t+z}{t-z}\biggr) \;.
\end{equation}
The temporal gauge in this coordinate, $\calA_\tau=0$, has a close
connection to the light-cone gauge discussed in the previous
subsection because $\calA_\tau = x^- \calA^+ + x^+ \calA^-=0$ leads to a
condition $\calA^+=0$ on $x^+=0$ and $\calA^-=0$ on $x^-=0$.
Therefore, the experience in the one-source problem turns out to be
useful here.

  The equations of motion can be expressed in the Bjorken coordinates
as
\begin{equation}
 \begin{split}
 \partial_\tau \calE^i &= \frac{1}{\tau}\calD_\eta \calF_{\eta i}
  +\tau\calD_j \calF_{ji} \;,\\
 \partial_\tau \calE^\eta &= \frac{1}{\tau}\calD_j \calF_{j\eta} \;.
 \end{split}
\label{eq:eom3}
\end{equation}
and the conjugate momenta are defined as
\begin{equation}
 \calE^i = \tau\partial_\tau\calA_i \;,\qquad
 \calE^\eta = \frac{1}{\tau}\partial_\tau\calA_\eta \;.
\label{eq:momenta}
\end{equation}
These equations~(\ref{eq:eom3}) and (\ref{eq:momenta}) determine the
time evolution uniquely once the initial condition at an initial time
$\tau_0$ is specified.

\begin{figure}
\begin{center}
\includegraphics[width=0.5\textwidth]{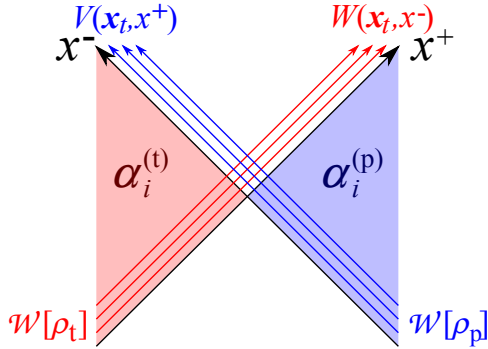}
\end{center}
\caption{Solution of the equations of motion~(\ref{eq:eom2}) outside
  of the forward light-cone.}
\label{fig:light-cone}
\end{figure}

  The solutions of the one-source problems are consistent with the
gauge-fixing condition $\calA_\tau=0$ since we found
$\calA^+=\calA^-=0$ for the target and the projectile both.  Then
$\alpha_i^{\rm (t)}$ solves Eq.~(\ref{eq:eom3}) too if there is no
interference from the projectile source.  This means that
$\alpha_i^{\rm (t)}$ should be a solution in the region outside of the
light-cone, $(x^+<0,\; x^->0)$, and $\alpha_i^{\rm (p)}$
in $(x^+>0,\; x^-<0)$ (see Fig.~\ref{fig:light-cone} for
illustration).

\begin{figure}
\begin{center}
\includegraphics[width=0.5\textwidth]{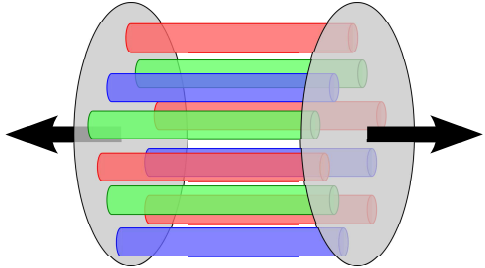}
\end{center}
\caption{Intuitive picture of the initial condition for the heavy-ion
  collision characterized by the CGC initial
  conditions~(\ref{eq:eom3}) and (\ref{eq:momenta}).}
\label{fig:glasma}
\end{figure}

  From this fact, it is naturally understood that the initial
condition at $\tau=0$ or $x^+=x^-=0$ may be a superposition of these
two solutions,
\begin{equation}
 \calA_i = \alpha_i^{\rm (t)} + \alpha_i^{\rm (p)}\;,\qquad
  \calA_\eta = 0 \;.
\label{eq:init}
\end{equation}
Here, though this is a very simple Ansatz by a superposition, the
consequence is quite non-trivial.  The field strength associated with
these gauge fields is
\begin{equation}
 \calB^i = 0 \;,\qquad
 \calB^\eta = \calF_{12} = -\rmi g \Bigl(
  \bigl[\alpha_1^{\rm (t)}, \alpha_2^{\rm (p)}\bigr]
  +\bigl[\alpha_1^{\rm (p)}, \alpha_2^{\rm (t)}\bigr]\Bigr) \;,
\end{equation}
and thus, the longitudinal component of the chromo-magnetic fields
appears from the non-Abelian interactions.  By solving the equations
of motion, one can also find similar expressions for the
chromo-electric fields~\cite{Kovner:1995ja}
\begin{equation}
 \calE^i = 0 \;,\qquad
 \calE^\eta = \rmi g \Bigl(\bigl[\alpha_1^{\rm (t)},
  \alpha_1^{\rm (p)}\bigr] + \bigl[\alpha_2^{\rm (t)},
  \alpha_2^{\rm (p)}\bigr]\Bigr) \;.
\label{eq:init-e}
\end{equation}
These field strengths stand for characteristic properties of the
initial condition of the relativistic heavy-ion collisions in the CGC
or the so-called Glasma picture, an intuitive illustration for which
is displayed in Fig.~\ref{fig:glasma}.  It is important to point out
that the initial conditions at $\tau=0$ are independent of $\eta$,
namely, boost invariant.  Because there is no explicit $\eta$ in the
equations of motion, boost invariance is kept during the time
evolution.

\subsection{McLerran-Venugopalan model}

  One of the simplest and reasonable Ans\"{a}tze for the wave function
is the Gaussian approximation, that is given by
\begin{equation}
 \calW_x[\rho] = \exp\biggl[ -\int\rmd^2\xt\rmd z
  \frac{|\rho(\xt,z)|^2}{2g^2\mu(z)^2}\biggr] \;,
\label{eq:Gauss}
\end{equation}
which defines the McLerran-Venugopalan (MV) model.  Here $z$
represents either $x^+$ or $x^-$.  In this model setup $\mu(z)$
characterizes the typical energy scale.  Indeed, once the parton
saturation manifests itself, any details in the structure are lost and
only the transverse parton density $\sim Q_s^2$ should be a relevant
scale.  In principle $\mu(z)$ in the MV model is to be interpreted as
$Q_s$ in the parton saturation.  With the Gaussian wave
function~(\ref{eq:Gauss}), the expectation value is obtained by a
decomposition into the two-point function that is read as
\begin{equation}
 \bigl\langle\rho^{(m)a}(\xt,z)\,\rho^{(n)b}(\yt,z')\bigr\rangle
  = g^2\mu^2(z) \,\delta^{mn}\,\delta^{ab}\,\delta(z-z')\,
  \delta^{(2)}(\xt-\yt) \;.
\end{equation}
In other words, the Gaussian approximation~(\ref{eq:Gauss}) assumes no
correlation at all between spatially distinct sites.  Evaluating the
Gaussian average with various functionals of $\rho^a(\xt,z)$ is an
interesting mathematical excercise~\cite{Fukushima:2007dy}.

Especially it is feasible to evaluate the initial energy density,
\begin{equation}
 \varepsilon = \bigl\langle T^{\tau\tau}\bigr\rangle =
  \bigl\langle\tr\bigl[ E_{_L}^2 + B_{_L}^2 + E_{_T}^2 + B_{_T}^2
  \bigr]\bigr\rangle
\end{equation}
at $\tau=0$ using the initial conditions~(\ref{eq:init}) and
(\ref{eq:init-e}) and the Gaussian weight~(\ref{eq:Gauss}).  The
results are a bit disappointing because it involves both the UV and
the IR divergences, which can be regularized by $\Lambda$ (momentum
cutoff) and $m$ (gluon mass).  Then, the initial energy density is
found to be
\begin{equation}
 \varepsilon(\tau=0) = g^6\mu^4 \cdot \frac{\Nc(\Nc^2-1)}{8\pi^2}\biggl(
  \ln\frac{\Lambda}{m}\biggr)^2
\label{eq:energy}
\end{equation}
after some
calculations~\cite{Fukushima:2007ja,Lappi:2006hq,Fujii:2008km}.  In
the numerical simulation with discretized grid in a finite-volume
box~\cite{Krasnitz:1998ns,Krasnitz:2001qu} there are natural UV and IR
cutoffs incorporated from the beginning.  The IR cutoff is, however,
not included as a gluon mass but originates from the system size $L$.

\section{Numerical method and the boost-invariant results}
\label{sec:numerical}

  The model parameters should be fixed first.  In the case of the
gold-gold collision at $\sqrt{s_{_{NN}}}=200\;\text{GeV}$, the
empirical choice is $g=2$ and $g^2\mu L = 120$, which correspond to
$\alpha\simeq 0.3$, $R_A\simeq 7\;\text{fm}$, and
$g^2\mu\simeq 2\;\text{GeV}$.  Now that the model definition and the
model parameters are given, we can calculate $\alpha_i^{\rm (t)}$ and
$\alpha_i^{\rm (p)}$ numerically.  Then, in the high-energy limit, it
is a common assumption that the nucleus source is infinitesimally
thin, i.e.
\begin{equation}
 \rhot(\xt,x^-) = \bar{\rho}_{\rm t}(\xt)\,\delta(x^-) \;,\qquad
 \rhop(\xt,x^+) = \bar{\rho}_{\rm p}(\xt)\,\delta(x^+) \;.
\end{equation}
Then, the Wilson lines are replaced, respectively, by
\begin{equation}
 \bar{V}^\dagger(\xt) \;\to\;
  \rme^{\rmi g \Lambda^{\rm (t)}(\xt)}\;,\qquad
 \bar{W}^\dagger(\xt) \;\to\;
  \rme^{\rmi g \Lambda^{\rm (p)}(\xt)}
\end{equation}
with the solution of the Poisson equation,
\begin{equation}
 -\bdel^2 \Lambda^{\rm (t)}(\xt) = \bar{\rho}_{\rm t}(\xt)\;,\qquad
 -\bdel^2 \Lambda^{\rm (p)}(\xt) = \bar{\rho}_{\rm p}(\xt)\;.
\end{equation}
One should be very careful about this replacement because this is not
an approximation on Eqs.~(\ref{eq:V}) and (\ref{eq:W}).  Even though
the longitudinal extent in the color source is infinitesimally thin,
it should not be legitimate to drop the path
ordering~\cite{Fukushima:2007ki}.  Therefore, we have to think that
the numerical MV model is something distinct from the original MV
model in the continuum variables.

\begin{figure}
\includegraphics[width=0.5\textwidth]{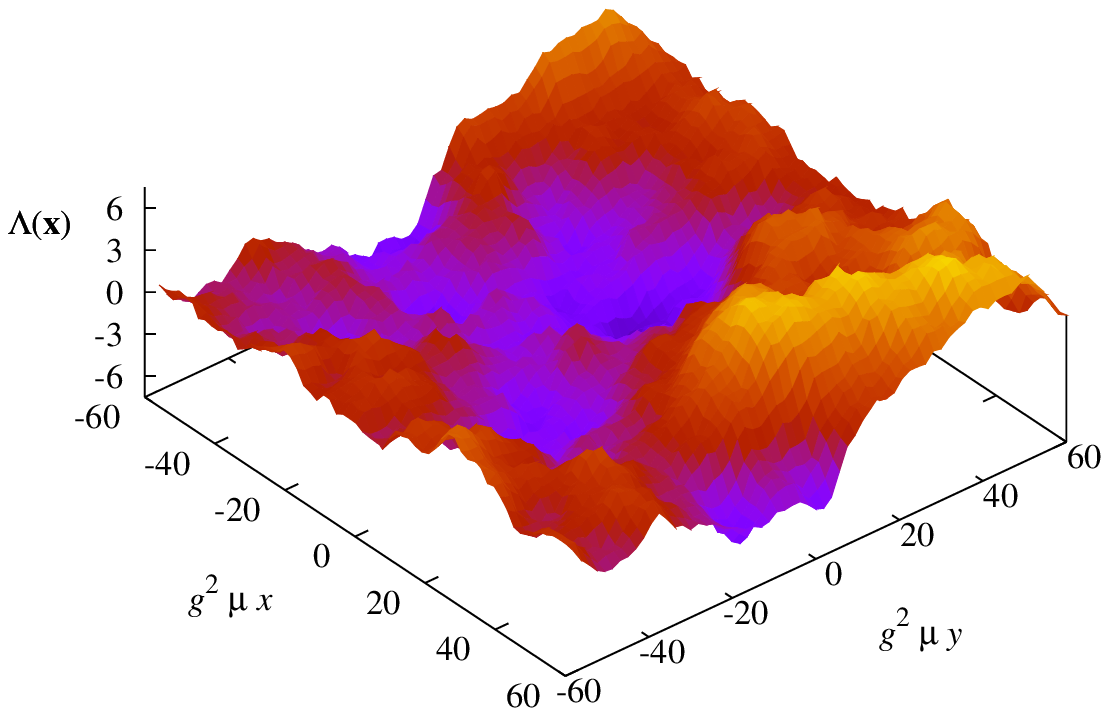}
\includegraphics[width=0.5\textwidth]{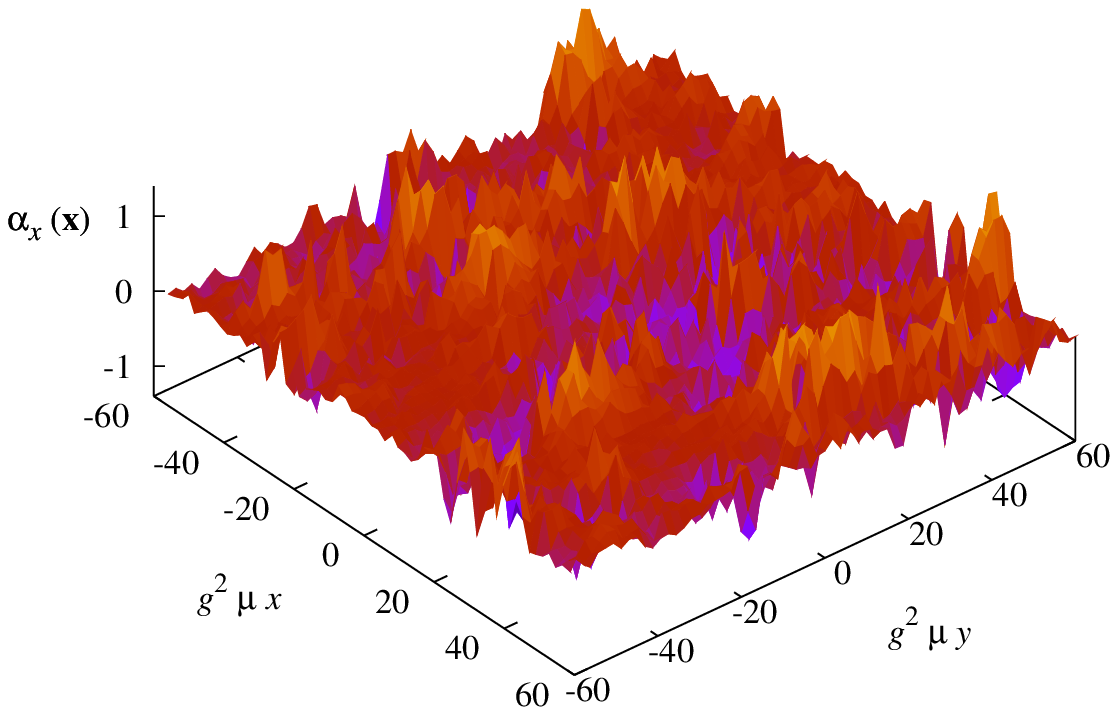}
\caption{Left: An example of the initial distribution of
  $\Lambda(\xt)$ as a solution of the Poisson equation with random
  color source.\ \ Right: The corresponding gauge field
  $\alpha_i=-(1/\rmi g)V\partial_i V^\dagger$.}
\label{fig:initial}
\end{figure}

  In the MV model the distribution of the color source is random and
there is no correlation between different sites.
Figure~\ref{fig:initial} illustrates an example of the initial
$\Lambda(\xt)$ as a solution of the Poisson equation.  Because the
operator $1/\bdel^2$ involves spatial average, we see that the spatial
distribution of $\Lambda(\xt)$ is rather smooth even though the source
$\bar{\rho}(\xt)$ has random fluctuations.  This smoothness is not
physical, however, and the gauge fields are furiously fluctuating as
shown in the right panel of Fig.~\ref{fig:initial}.  It is worth
noting that the color-flux tube picture as sketched in
Fig.~\ref{fig:glasma} is not the case in the MV model and the JIMWLK
evolution is indispensable to take account of the flux tube
structure.

  The physical observables are measured by taking an ensemble average
of results with different initial $\bar{\rho}(\xt)$'s.  It is useful
to compute not only the energy density~(\ref{eq:energy}) but also
other combinations of the energy-momentum tensor.  In particular the
following pressures are important in order to judge how anisotropic
the system is;
\begin{align}
 &P_{_T} = \frac{1}{2}\bigl\langle T^{xx}+T^{yy}\bigr\rangle
  = \bigl\langle\tr\bigl[ E_{_L}^2+B_{_L}^2\bigr]\bigr\rangle\;,\\
 &P_{_L} = \bigl\langle \tau^2 T^{\eta\eta}\bigr\rangle
  =\bigl\langle\tr\bigl[ E_{_T}^2+B_{_T}^2-E_{_L}^2-B_{_L}^2
  \bigr]\bigr\rangle\;,
\end{align}
where the longitudinal and transverse chromo-electric and
chromo-magnetic fields are defined as
\begin{align}
 &E_{_L}^2 = \calE^{\eta a}\calE^{\eta a}\;,\qquad
  E_{_T}^2 = \frac{1}{\tau^2}\bigl(\calE^{xa}\calE^{xa}
  +\calE^{ya}\calE^{ya}\bigr)\;,\\
 &B_{_L}^2 = \calF_{12}^a \calF_{12}^a\;,\qquad
  B_{_T}^2 = \frac{1}{\tau^2}\bigl( \calF_{\eta x}^a \calF_{\eta x}^a
  +\calF_{\eta y}^a \calF_{\eta y}^a \bigr) \;.
\end{align} 

\begin{figure}
\includegraphics[width=0.5\textwidth]{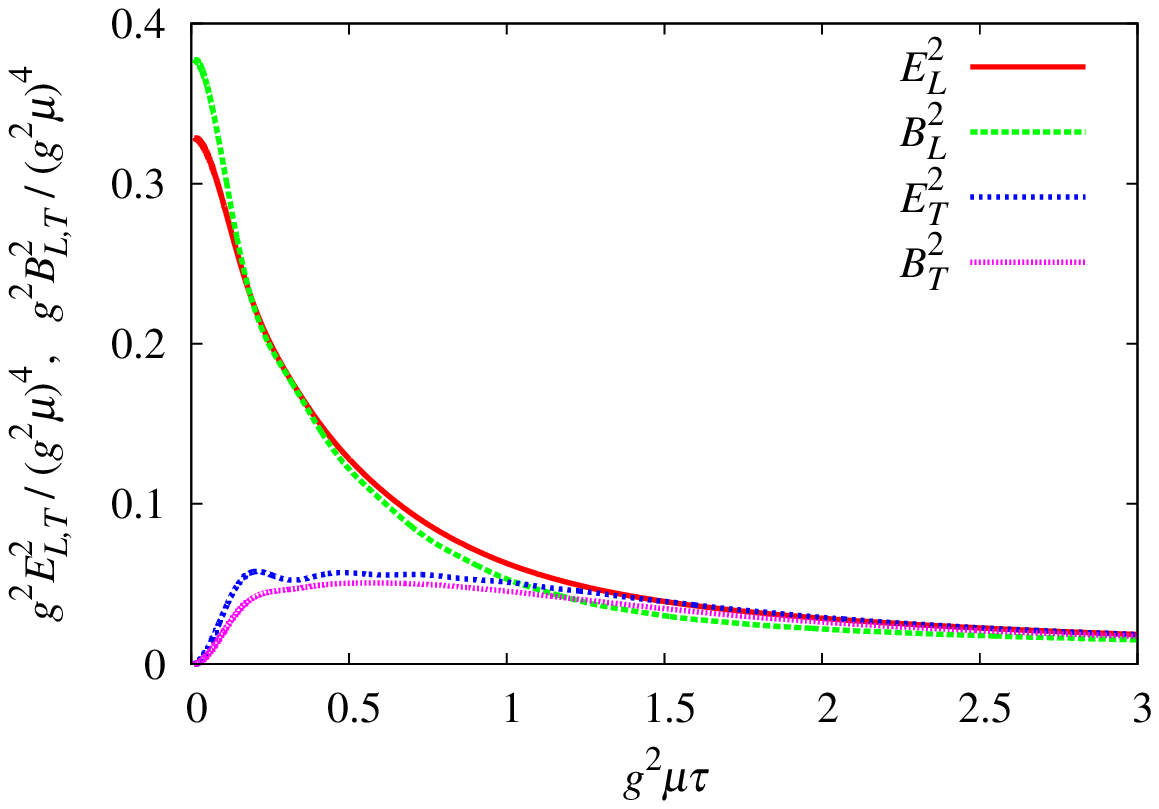}
\includegraphics[width=0.5\textwidth]{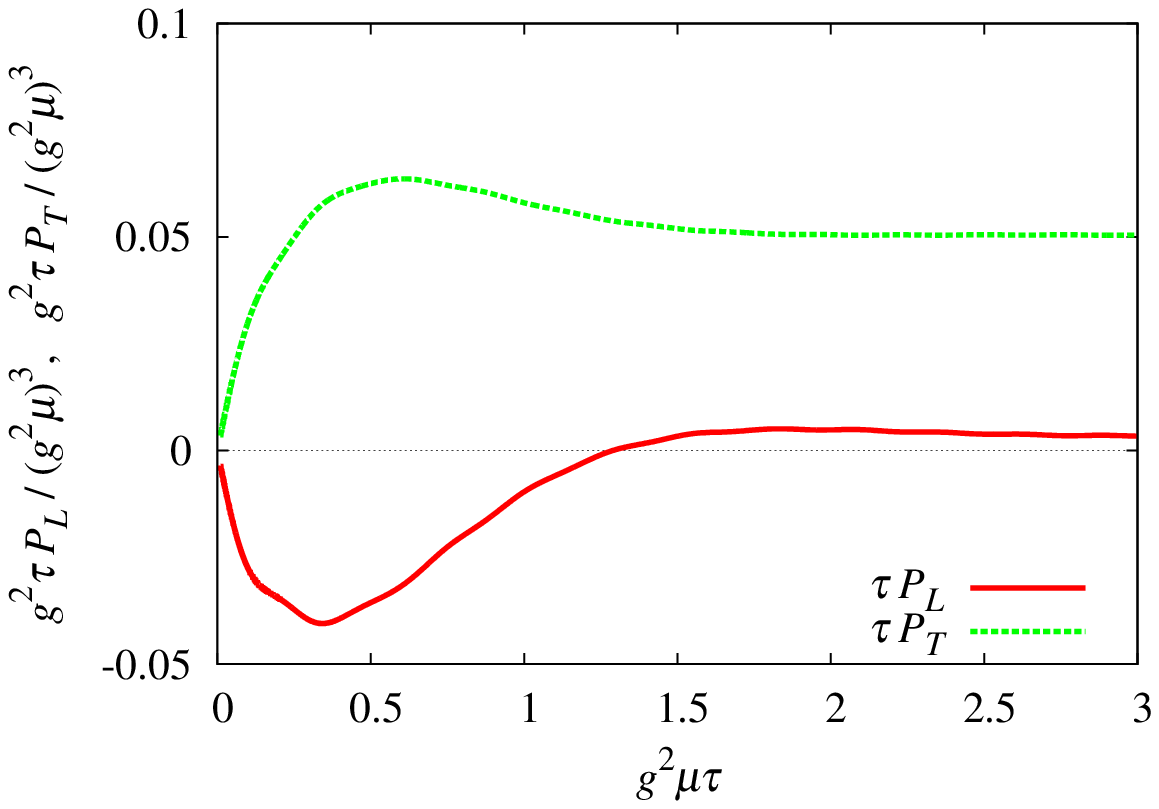}
\caption{Left: Time evolution of the chromo-electric and the
  chromo-magnetic fields.  The subscripts, $L$ and $T$, represent the
  longitudinal and the transverse fields, respectively.\ \ Right: The
  longitudinal and the transverse pressures as a function of time.}
\label{fig:EB}
\end{figure}

  The numerical results from the numerical Glasma simulation are
presented in Fig.~\ref{fig:EB}.  From this figure it is clear that
there are only longitudinal fields $E_{_L}^2$ and $B_{_L}^2$ right at
the collision ($\tau=0$) as explained with Fig.~\ref{fig:glasma}.  The
transverse fields are developing as $\tau$ increases, and eventually
the longitudinal and the transverse fields approach each other at
$g^2\mu\tau>1$.  This does not mean isotropization, however, because
there are two ($x$ and $y$) components in the transverse direction.
One can understand what is happening by taking a careful look at the
pressures.  The longitudinal pressure $P_{_L}$ goes to zero for
$g^2\mu\tau>1$ and this means that particles move on the free stream
along the longitudinal direction.  Thus, the system remains far from
thermal equilibrium.

  Here we encounter a perplexing situation.  We know the the CGC
should be a correct description of the initial dynamics of the
heavy-ion collision.  Even if the CGC cannot reach isotropization and
thermal equilibrium, it should be a natural anticipation that the CGC
can at least capture a correct tendency toward thermalization.  This
anticipation is not the case at all, however, as seen in
Fig.~\ref{fig:EB}.  Then, is there anything missing in our
discussions so far?

\section{Glasma instability and the scaling spectrum}
\label{sec:inst}

  Yes, there is.  We have neglected fluctuations on top of
boost-invariant CGC-background field and thus $\eta$ dependence at
all.  Such a treatment is not always justified.  As a matter of fact,
longitudinal structures in a finite extent by not using
$\bar{\rho}_{\rm t}(\xt)\,\delta(x^-)$ but treating $\rhot(\xt,x^-)$
correctly in the path ordering would give rise to $\eta$-dependent
fluctuations.  Also, quantum fluctuations have $\eta$
dependence as well~\cite{Dusling:2011rz,Fukushima:2006ax}.

\begin{figure}
\includegraphics[width=0.5\textwidth]{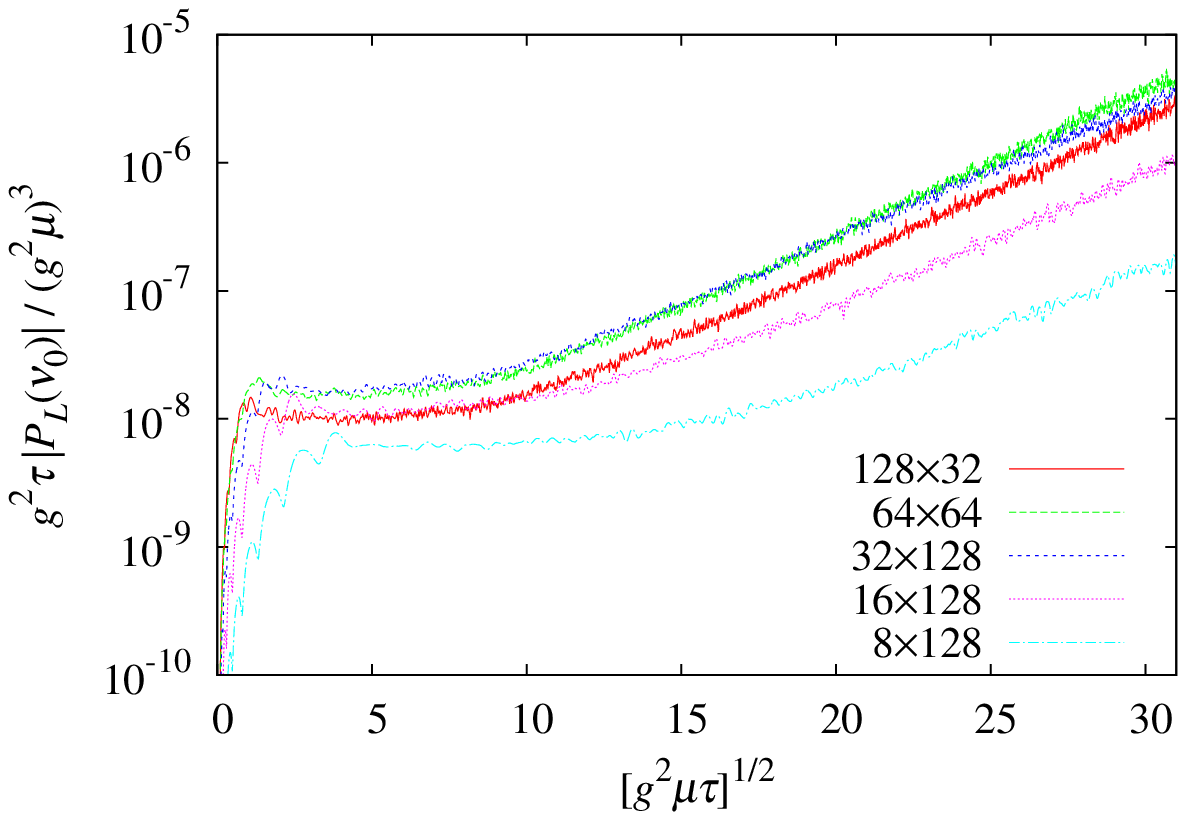}
\includegraphics[width=0.5\textwidth]{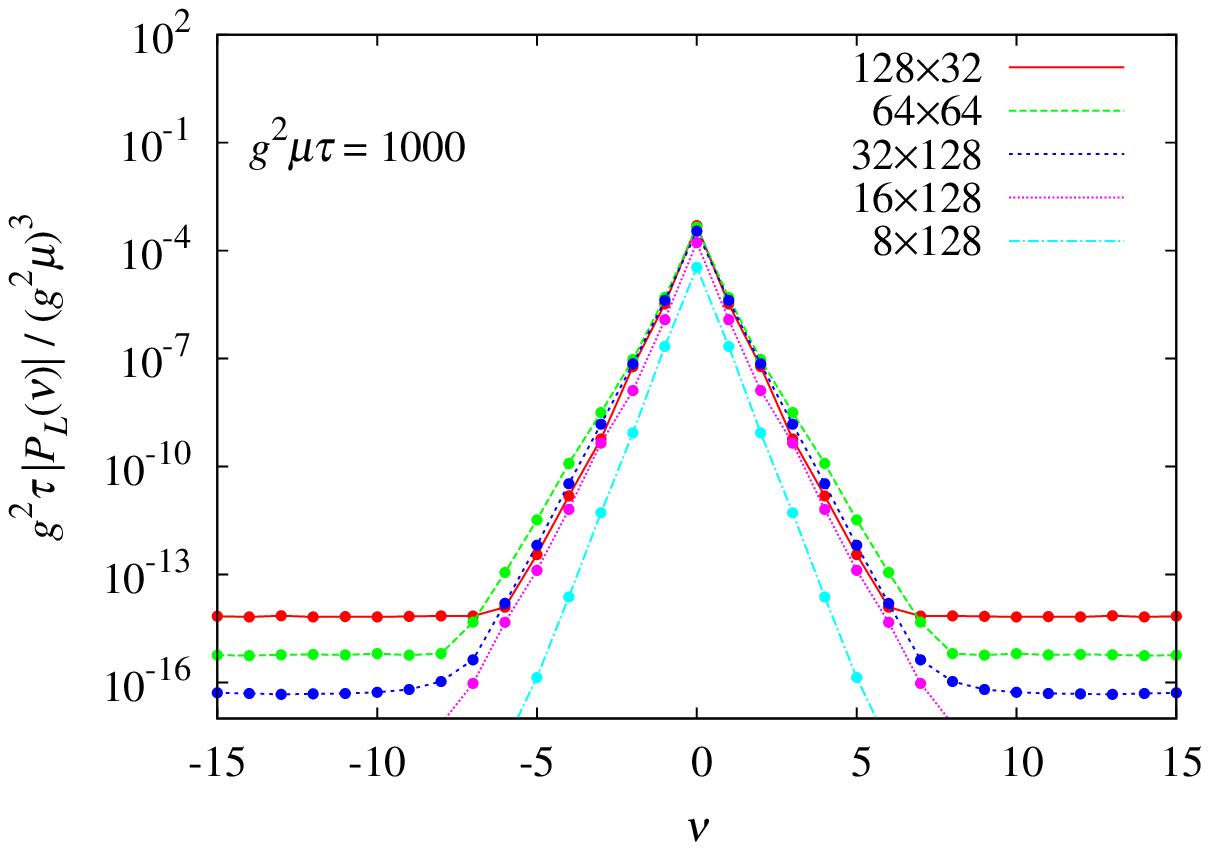}
\caption{Left: Longitudinal pressure at $\nu=\nu_0=1$ for various
  system sizes.\ \ Right: Spectrum of the longitudinal pressure at
  late time presented in $\nu$ space.}
\label{fig:inst}
\end{figure}

  One might have thought that small disturbances in the longitudinal
direction could make only a slight difference.  But, the fact is that
there is a tremendous difference between results with and without
$\eta$-dependent fluctuations.  With random fluctuations in $\eta$
space, the Glasma simulation would lead to significant decay from the
CGC background fields with $\eta$-independent zero-mode into
$\eta$-fluctuating non-zero modes~\cite{Romatschke:2005pm}.  An
example is shown in Fig.~\ref{fig:inst}.

To obtain the results as presented in Fig.~\ref{fig:inst}, only
$\nu=\pm\nu_0$ modes are disturbed,
\begin{equation}
 \delta E^\eta \;\propto\; f(\eta) = \Delta \cos\biggl(
  \frac{2\pi\nu_0}{L_\eta}\eta\biggr)\;,
\label{eq:feta}
\end{equation}
where $L_\eta$ is the longitudinal extent that we took as $L_\eta=2$
in the simulation.  Once $\delta E^\eta$ is given, the transverse
fluctuations, $\delta E^i$, are chosen in such a way to satisfy the
Gauss law.  Then, physical observables of our interest are Fourier
transformed from $\eta$ space to $\nu$ space.  It is now obvious that
the CGC background fields at $\nu=0$ are relatively larger than
fluctuations by small $\Delta$ and the second dominant mode should sit
at $\nu=\nu_0$.  In this article we limit ourselves to the simplest
choice of $\nu_0=1$.

  Usually unstable modes grow up exponentially, but in the expanding
geometry, the instability implies a slightly weaker growth according
to $\sim \exp[\alpha \sqrt{\tau}]$.  The horizontal axis in the left
panel of Fig.~\ref{fig:inst} is, thus, not the dimensionless time
$g^2\mu\tau$ itself, but $\sqrt{g^2\mu\tau}$.  We can surely confirm
that the longitudinal pressure component at $\nu=\nu_0=1$ increases
linearly in Fig.~\ref{fig:inst} when plotted with the logarithm of the
pressure as a function of $\sqrt{g^2\mu\tau}$.

  The detailed structure of the instability in the spectral pattern is
interesting to see.  The right panel of Fig.~\ref{fig:inst} is the
spectrum corresponding to the simulation results shown in the left.

\begin{figure}
\includegraphics[width=0.5\textwidth]{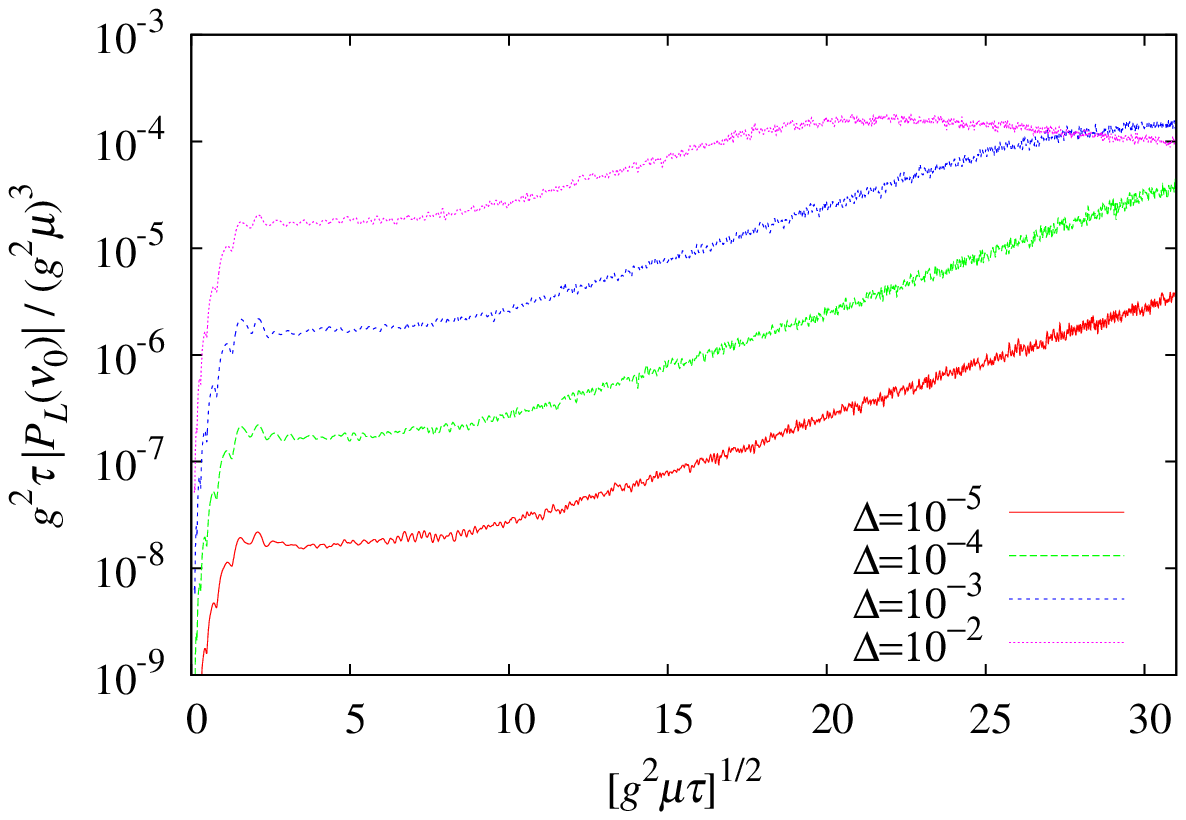}
\includegraphics[width=0.5\textwidth]{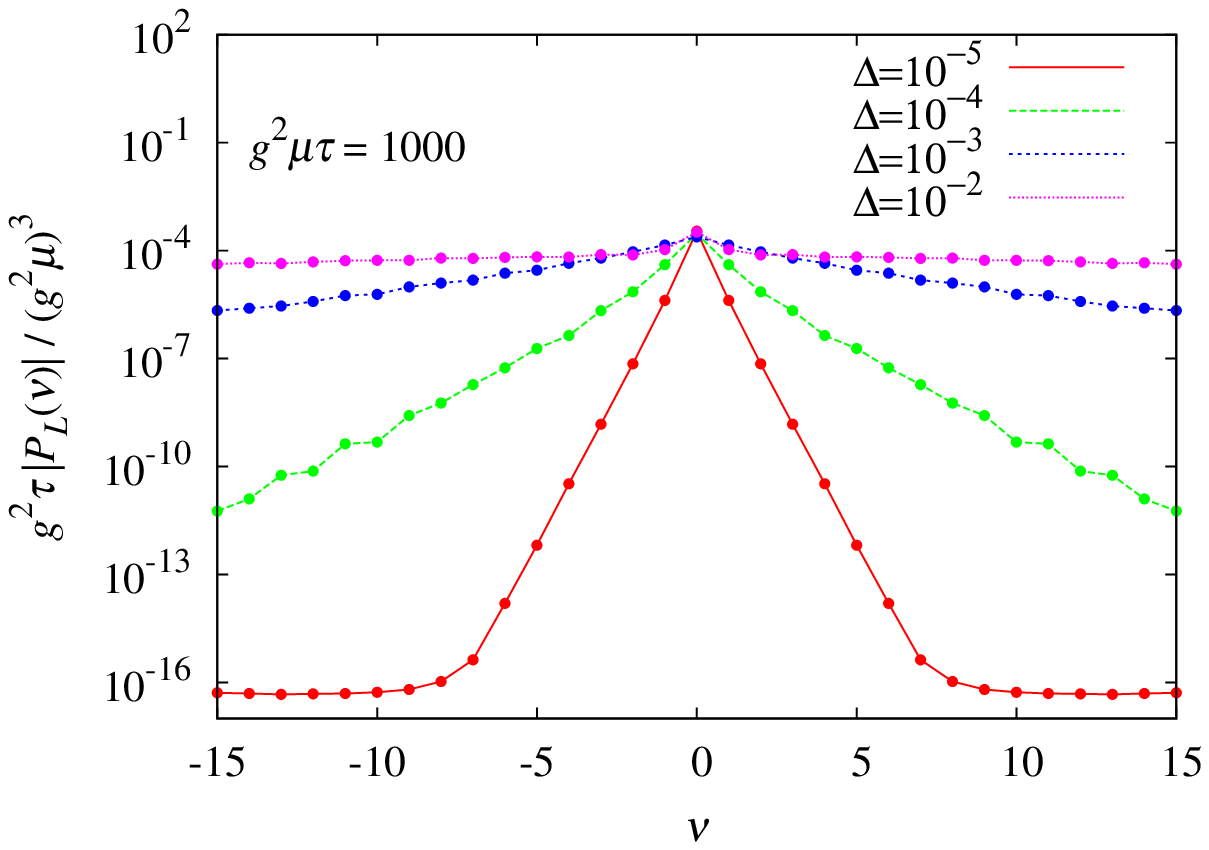}
\caption{Left: Instability as a function of the magnitude of the
  seed.  Saturation behavior in the non-linear regime is seen slightly
  for $\Delta=10^{-3}$ and clearly for $\Delta=10^{-2}$.\ \ Right:
  Corresponding spectra in the $\nu$ space.}
\label{fig:nonlinear}
\end{figure}

  From now on, let us consider the fate of the instability;  it is
simply impossible for the instability to last for ever.  The
saturation of instability growth can be observed in two ways.  The
first case is that the initial magnitude $\Delta$ (appearing in
Eq.~(\ref{eq:feta})) is taken to be large enough to accommodate
non-linear effects.  The second is the large-time behavior -- simply
we wait until the unstable modes grow up considerably.

  As shown in the left panel in Fig.~\ref{fig:nonlinear} the
instability for $\Delta=10^{-2}$ stops and the spectrum is flattened
after all.  In the saturated regime at $\sqrt{g^2\mu\tau}>20$ we see
that the pressure at $\nu=\nu_0$ slightly decreases rather than
increasing.  This behavior can be interpreted as follows;  As long as
the non-zero modes are small (i.e.\ in the linear regime), the energy
decay from the dominant CGC background at $\nu=0$ makes non-zero modes
enhanced exponentially.  The injected energy is much bigger than the
escaped energy toward higher-$\nu$ modes in the linear regime.  This
balance changes gradually with increasing amplitude of unstable modes,
and eventually a steady energy flow is expected when non-Abelian
nature of non-zero modes becomes substantially large (i.e.\ in the
non-linear regime).  At even larger time, as hinted from
Fig.~\ref{fig:nonlinear}, the spectrum is flattened and the UV-cutoff
effects at large $\nu$ should be appreciable.  Then, an intriguing
question is;  what is going on in the non-linear regime before the
UV-cutoff effects contaminate the simulation?

\begin{figure}
\begin{center}
\includegraphics[width=0.5\textwidth]{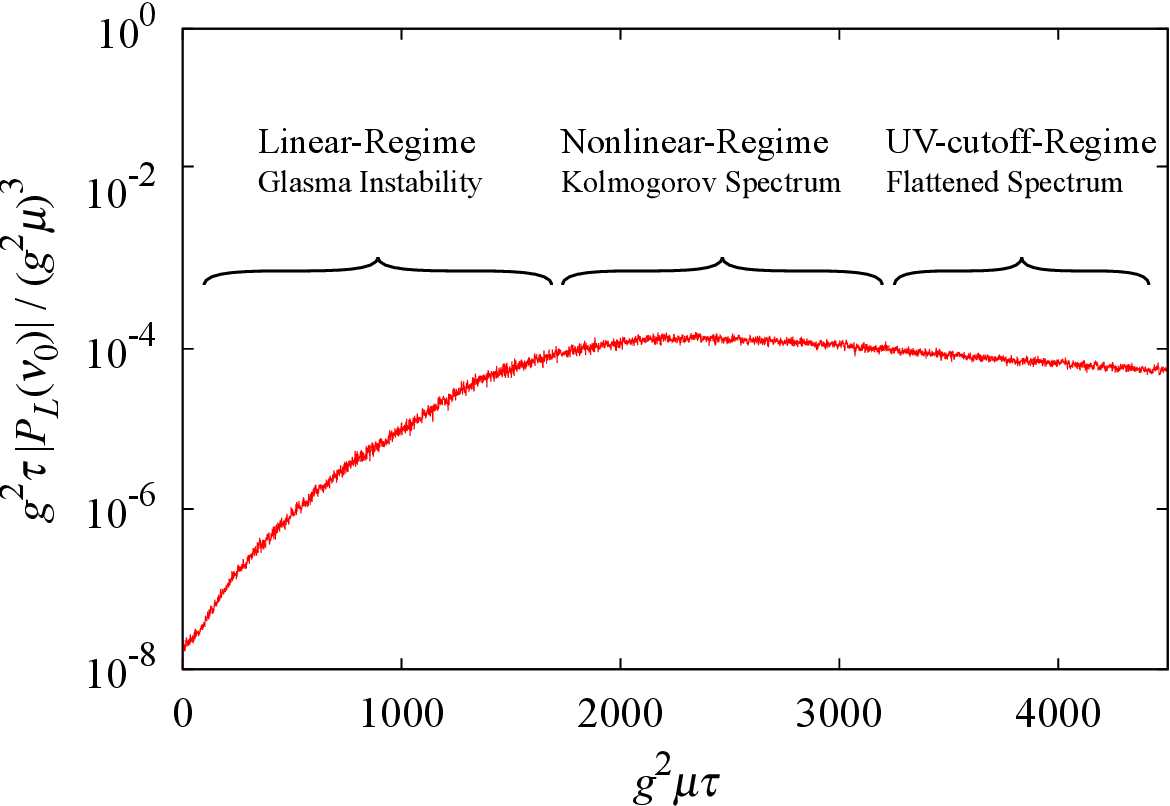}
\end{center}
\caption{Saturation of the instability at late time with indications
  to three distinct regions;  the linear regime where the instability
  persist, the non-linear regime where the energy flow is expected,
  and the UV-cutoff regime where artificial results by the UV cutoff
  are unavoidable.}
\label{fig:long}
\end{figure}

  To address this question, we shall take the latter case, namely the
long-time run of the simulation with small $\Delta$ for better
numerical stability.  The qualitative features in the results for
$\Delta=10^{-5}$ shown in
Fig.~\ref{fig:long} are just the same as the top curve in
Fig.~\ref{fig:nonlinear} that represents the results for
$\Delta=10^{-2}$.  We are interested in the energy spectrum in the
non-linear regime that can be immediately identified on
Fig.~\ref{fig:long}.

\begin{figure}
\includegraphics[width=0.5\textwidth]{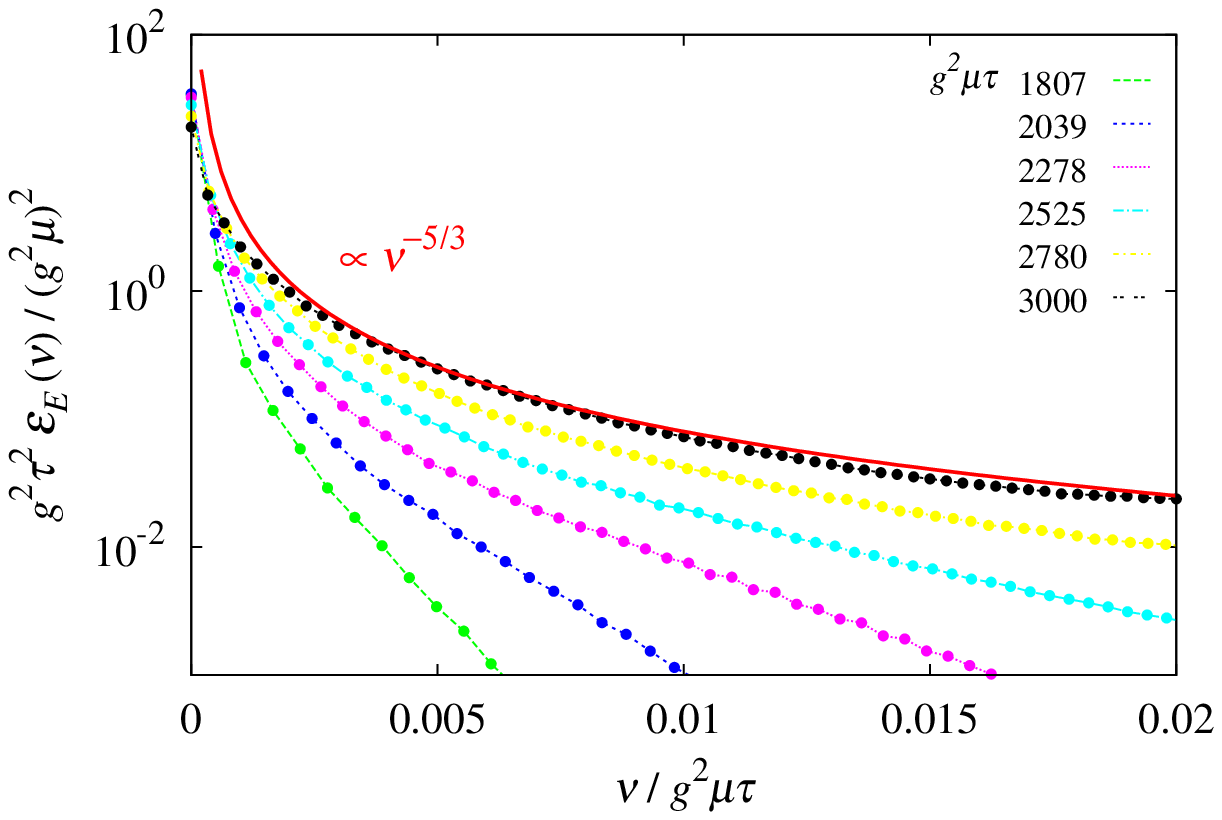}
\includegraphics[width=0.5\textwidth]{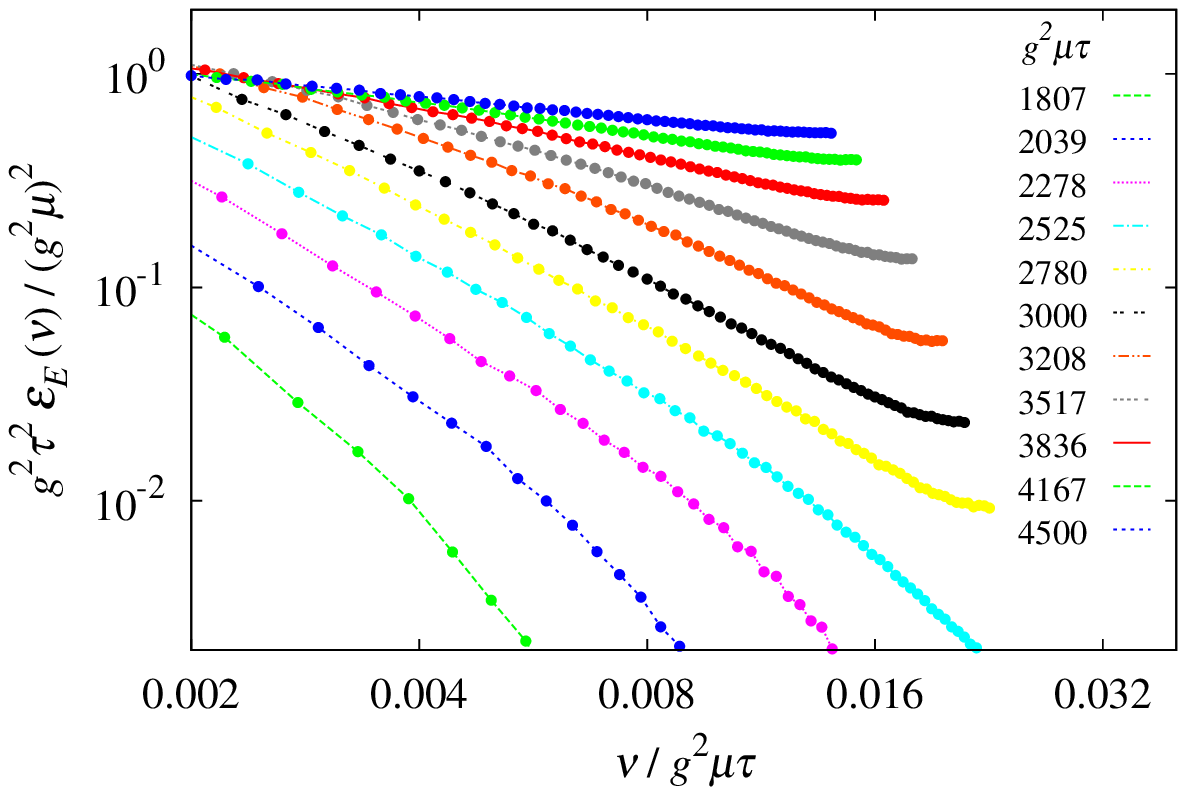}
\caption{Left: Energy spectrum approaching the power-law scaling
$\propto \nu^{-5/3}$.\ \  Right: Breakdown of the Kolmogorov-type
  scaling in the UV-cutoff regime.}
\label{fig:scaling}
\end{figure}

  Figure~\ref{fig:scaling} is the energy spectrum (not the
Fourier-transformed longitudinal pressure but the energy contribution
from respective $\nu$ modes).  That is, what is shown in
Fig.~\ref{fig:scaling} is given as
\begin{equation}
 \varepsilon_{_E}(\nu) = \bigl\langle\tr\bigl[ \calE^{\eta a}(-\nu)
  \calE^{\eta a}(\nu) + \tau^{-2} \calE^{i a}(-\nu)\calE^{i a}(\nu)
  \bigr]\bigr\rangle \;.
\end{equation}
We see that there is a clear tendency to approach a scaling form in
the non-linear regime (as in the right panel of
Fig.~\ref{fig:scaling}).  With some rescaling we find that this power
is exactly consistent with the Kolmogorov value, $-5/3$.  Generally
speaking, in non-expanding systems, it is not a surprise that the
Kolmogorov power emerges because this value can be guessed by
dimensional analysis.  Such a dimensional argument may work even for
the expanding system.  In the Bjorken coordinates $\eta$ is
dimensionless, but the physical scale in the longitudinal direction is
to be interpreted as $\tau\eta$, and thus the corresponding momentum
should be $\nu/\tau$.  Then, this quantity gives a dimension of
length (with $c$ multiplied appropriately).  Because of an expected
energy flow in $\nu$ space, its rate $\psi$ is also a relevant
quantity.  Then, the characteristic length and time scales of the
system are expressed by two quantities with the following dimensions;
\begin{equation}
 [\nu/\tau] = l^{-1}\;,\qquad
 [\psi] = l^2\,t^{-3}\;.
\end{equation}
The energy spectrum, on the other hand, has the dimension,
\begin{equation}
 \bigl[V_\perp \tau^2\,\varepsilon_{_E}(\nu)\bigr] = l^3\,t^{-2} \;,
\end{equation}
that is reproduced exactly by a unique combination of
$(\nu/\tau)^{-5/3}(\psi)^{2/3}$.  Therefore, it is concluded that
$\tau^2\varepsilon_{_E}(\nu/\tau)$ should exhibit the power-law
scaling in terms of $\nu/\tau$ whose power is identical to the
Kolmogorov's value, $-5/3$ as long as the system stays in the
non-linear regime.  From the left panel of Fig.~\ref{fig:scaling}, we
can see only the scaling region or the so-called inertial region
realized.  The dissipative region at high $\nu$ is not found, probably
because of the UV-cutoff effects.

  This nice scaling is lost at further later time.  We can understand
from the right panel of Fig.~\ref{fig:scaling} that the energy flow is
stuck at the UV edge and the energy spectrum is artificially pushed up
by the UV-cutoff contamination.

\section{Outlooks}
\label{sec:outlook}

  It was a surprise that the Kolmogorov's scaling law could be
realized in the Glasma simulation in the expanding geometry.  The
dimensional argument is not so strong to constrain the shape of the
energy spectrum uniquely.  It should be an important test whether
the Kolmogorov's $-5/3$ behavior can be confirmed or not in other
simulations of the pure Yang-Mills theory.  It has been established
that the non-Abelian plasma generally has an instability associated
with anisotropy that grows up until the non-Abelianization
occurs~\cite{Arnold:2005vb}.  Similar phenomena of instability tamed
by non-Abelian interactions are found in many other examples.  Then,
presumably, there must appear the power-law scaling in the region
around the non-Abelianization.

  The turbulent decay and the associated Kolmogorov's power-law are
interesting discoveries from the long-run simulation of the Glasma.
But, it cannot answer anything about the realistic thermalization
mechanism from the Glasma.  The turbulence is certainly a tendency
into thermalized matter, but the energy flow is a slow process and the
relevant time scale, $g^2\mu\tau\sim 2000$, is far outside of the
validity region of the CGC description.  There must be still something
missing that can accelerate the thermalization speed.

  If this missing piece were finally set in, the transient Glasma
would provide us with the initial input for the hydrodynamic
equations.  Even in this case, the analysis we have seen should be
useful.  The energy spectrum in $\nu$ space should carry important
information.  Then, the power could agree with or deviate from $-5/3$
as suggested in strong-coupling
studies~\cite{Berges:2008zt,Carrington:2010sz}.

\section*{Acknowledgments}
The author thanks the organizers of the 51st Cracow School of
Theoretical Physics for a kind invitation.  This article summarizes
lectures given there.  He also thanks Hiro~Fujii, Francois~Gelis,
and Yoshimasa~Hidaka for collaborations.  The central parts in these
lectures are based on works done in collaborations with them.

\end{document}